\documentstyle[12pt]{article}
\begin{document}

In a recent letter to Nature, Edwards and Ashcroft\cite{ea} made the 
interesting suggestion that phase III of solid molecular hydrogen be 
due to an electronic instability arising at a density 
${\rho\over\rho_\circ}\simeq 9$, when the pressure is about 150~GPa. 

While we agree completely with the above authors that the dramatic 
increase in the infrared absorption together with the similar 
behaviours of $H_2$ and $D_2$ in phase III strongly
suggest that for P$>150$ GPa a new electronic configuration characterizes 
solid molecular hydrogen, we have strong reservations about their explanation 
in terms of a dielectric instability. And this for the following reasons:
even though we may accept that in a mean field approximation the electric 
field at the site of a molecule is $\vec {E}=\left({N\over V}\right)f\vec{d}$ 
(~$\vec{d}$~is the static electric dipole of the molecule), $f\simeq 5$, and 
the energy gain per molecule due to the electrostatic interaction  
$\Delta_{es}=-\left({N\over V}\right)f(\vec{d})^2$, 
we have good reasons to doubt
that the latter is sufficient to overcome the energy deficit necessary to bring 
the molecule to the electronic state of polarization that characterizes 
phase III.
 
Indeed, from the observed infrared absorbance one can determine
the experimental oscillator strength of the vibron mode \cite{mh} $f_{exp}
\simeq 1.2 10^{-5}$. By its definition ( we are using natural units, 
where $\hbar=c=1$)
\begin{equation}
e^{2}f=\frac{2}{3}m_{e}\omega\vec{d}^{2}
\end{equation}
where $e$ is the electron's charge, $m_e$ its mass and 
$\omega$ the vibron frequency. 
 Substituting the experimental numbers one thus gets,
\begin{equation}
|\vec{d}|\simeq 5\cdot 10^{-11}{\rm cm},
\label{d}
\end{equation}
which gives $\Delta_{es}\simeq -0.07$eV. On the other hand, the molecules in 
phase III, due to the still high
gap between the conduction and the valence band, can be well represented by
the quantum state ($|0\rangle$ is the ground state, whose electric dipole is
zero),
\begin{equation}
|III\rangle\simeq \cos\theta |0\rangle + \sin\theta |\Pi_u\rangle,
\label{m}
\end{equation}
$|\Pi_u \rangle$ being an excited state with energy $E(\Pi_u)=12.4$ 
eV
and dipole moment $|\vec{d}(\Pi_u)|\simeq 2\cdot 10^{-9}$cm. Thus we may write
\begin{equation}
|\vec{d}|\simeq 5\cdot 10^{-11}{\rm cm}=\sin^2\theta |\vec{d}(\Pi_u)|,
\label{dd}
\end{equation}
i.e.~$\sin^2\theta\simeq 0.08$. This implies that the energy deficit
$\Delta_{pol}=\sin^2\theta E(\Pi_u)$ for polarizing the hydrogen molecules in 
phase III is about 1 eV, much bigger than the estimated electrostatic gain 
of 0.07 eV.

In a recent paper \cite{bdpx} we have shown that no such difficulty arises when 
a peculiar electrodynamic instability is invoked and analysed quantitatively.

\vskip1cm
Yours Sincerely,
\begin{center}
M.~Buzzacchi, E.~Del Giudice, G.~ Preparata, S.S.~Xue 
\end{center}
\begin{center}
{\it Dipartimento di Fisica dell'Universit\`{a} di Milano}
\end{center}
\begin{center}
{\it Via Celoria 16 - 20133 Milano, Italy}
\end{center}

\end{document}